\documentclass[aps,prl,twocolumn,10pt,showpacs,superscriptaddress,amsmath,amssymb,nobibnotes]{revtex4-1}

\usepackage{graphicx}
\usepackage{amsfonts}    
\usepackage{graphicx}   
\usepackage{verbatim}   
\usepackage{color}      
\usepackage{subfigure}  
\usepackage{hyperref}   
\usepackage{natbib}
\usepackage{booktabs}
\usepackage{threeparttable}
\usepackage{dcolumn}
\usepackage{multirow}

\begin{document}
\title{Cascaded Weak Measurements Amplification of Ultra-small Phases}
\author{Meng-Jun Hu}\email{mengjun@mail.ustc.edu.cn}
\author{Yong-Sheng Zhang}\email{yshzhang@ustc.edu.cn}
\affiliation{Laboratory of Quantum Information, University of Science and Technology of China, Hefei 230026, China}
\affiliation{Synergetic Innovation Center of Quantum Information and Quantum Physics, University of Science and Technology of China, Hefei 230026, China}

\date{\today}

\begin{abstract}
The weak measurements based amplification of ultra-small phase was proposed in our previous work. Due to the technical imperfections, the ability of amplification is usually limited in practice. Here we introduce the concept of cascaded weak measurements amplification so that the ability of weak measurements amplification can be further improved. A formula of determining the cascaded number of weak measurements amplification is derived. The method presented here could be found useful in high precision measurement e.g., gravitational waves detection.
\end{abstract}

\maketitle
The amplification of ultra-small phases by using weak measurements has been discussed and a weak measurement amplification based laser interferometer gravitational-wave observatory i.e., WMA-LIGO was suggested in our previous work \cite{hu}. Here, we introduce the concept of cascaded weak measurements amplification of ultra-small phases with which the amplification factor limited by technical imperfections can be further improved.

The key idea of weak measurements based ultra-small phase amplification (WMPA), which is contrary to weak value amplification (WVA) \cite{wva1, wva2, wva3, wva4, wva5}, is to transform the ultra-small phase into the larger rotation along the equator of the Bloch sphere via appropriate post-selection \cite{hu}. The WMPA overcomes the difficulties of ultra-small phase amplification by using WVA and provides a universal amplification scheme of ultra-small phase. Contrary to WVA, which is a linear amplification with amplification factor determined by weak value, the WMPA is nonlinear, and this implies that WMPA opens a new way of amplification in the framework of weak measurements. 

To see how WMPA works, we consider a two-level system initially prepared in the superposition state of $|\psi_{i}\rangle_{S}=(|0\rangle+|1\rangle)/\sqrt{2}$. To simplify the calculation and practical realization, we choose a discrete pointer e.g., polarization of photons initially prepared in diagonal state $|+\rangle_{P}=(|H\rangle+|V\rangle)/\sqrt{2}$ with $|H\rangle$ and $|V\rangle$ represent horizontal and vertical polarization states respectively. The initial composite state of photons is $|\Psi\rangle_{SP}=|\psi_{i}\rangle_{S}\otimes|+\rangle_{P}$ with $|0\rangle,|1\rangle$ represent two different path states. The state of $|\Psi\rangle_{SP}$ can be easily prepared by sending photons in polarization state $|+\rangle$ through a beam splitter (BS). As for the interaction between the system and the pointer, we introduce the unitary operation of a control-rotation
\begin{equation}
\hat{U}=|0\rangle\langle 0|\otimes\hat{I}+|1\rangle\langle 1|\otimes(|H\rangle\langle H|+e^{i\theta}|V\rangle\langle V|)
\end{equation}
with $\theta$ is the ultra-small phase to be measured. The evolved state of photons becomes
\begin{equation}
|\Psi_{f}\rangle_{SP}=\dfrac{1}{2}[|0\rangle\otimes(|H\rangle+|V\rangle)+|1\rangle\otimes(|H\rangle+e^{i\theta}|V\rangle)].
\end{equation}
The control-rotation rotates the state of the pointer along the equator of the Bloch sphere with an angle $\theta$ if the system is in state $|1\rangle$ and does nothing for state $|0\rangle$. The control-rotation operation introduces phase signal into the pointer's state in a natural way and can be easily realized in practice e.g., via polarization-type Michelson interferometer \cite{hu, bird}. 
In order to realize amplification of the phase signal $\theta$, post-selection of the system after interaction is necessary. Suppose that the post-selected state has the form of $|\psi_{f}\rangle_{S}=\mathrm{cos}\chi|0\rangle+\mathrm{sin}\chi|1\rangle$, then the pointer's state after post-selection becomes (unnormalized)
\begin{equation}
|\tilde{\varphi}\rangle_{P}=_{S}\langle\psi_{f}|\Psi_{f}\rangle_{SP}=(\mathrm{cos}\chi+\mathrm{sin}\chi)|H\rangle+(\mathrm{cos}\chi+\mathrm{sin}\chi e^{i\theta})|V\rangle.
\end{equation}
In the polar coordinates $\mathrm{cos}\chi+\mathrm{sin}\chi e^{i\theta}=\sqrt{\mathrm{cos}^{2}\chi+\mathrm{sin}^{2}\chi+2\mathrm{sin}\chi\mathrm{cos}\chi\mathrm{cos}\theta}e^{i\gamma}$ with
$\mathrm{tan}\gamma=\mathrm{sin}\theta/(\mathrm{cos}\theta+\mathrm{cot}\chi)$. Since the phase $\theta\ll 1$, in the first order approximation $\mathrm{cos}\theta=1$ and $\mathrm{cos}\chi+\mathrm{sin}\chi e^{i\theta}=(\mathrm{cos}\chi+\mathrm{sin}\chi)e^{i\gamma}$. The normalized state of the pointer after post-selection thus has the form of
\begin{equation}
|\varphi\rangle_{P}=\dfrac{1}{\sqrt{2}}(|H\rangle+e^{i\gamma}|V\rangle)
\end{equation}
with error of order $o(\theta^{2})$. The amplification factor $h$, in the first order approximation, is determined by
\begin{equation}
h=\dfrac{\gamma}{\theta}=\dfrac{\mathrm{arctan}[\theta/(1+\mathrm{cot}\chi)]}{\theta},
\end{equation}
which illustrates that the amplification factor $h$ is determined solely by the choice of post-selected state $|\psi_{f}\rangle_{S}$. A large $h$ can be obtained by letting $\chi=-(\pi/4+\delta)$ with $\delta\ll 1$ represents the precision of $\chi$ such that $\mathrm{cot}\chi=-1+\delta$ in the first order approximation and
\begin{equation}
h=\dfrac{\mathrm{arctan}(\theta/\delta)}{\theta}.
\end{equation}
It should be noted here that $\delta=0$ is forbidden by the process of amplification though $h$ still has a definite value. In the case of $\delta=0$, the state of the pointer $|\tilde{\varphi}\rangle_{P}$ after post-selection will be $|V\rangle$ according to Eq. (3) and the phase signal $\theta$ is lost in a global phase that cannot be extracted in any way.

In practice, the angle $\delta$, which determines the ability of amplification, cannot be arbitrarily small due to technical imperfections e.g., non-ideal of optical elements and its calibration. There are two ways to improve the ability of amplification further, one depends on significant technical improvement such that high quality and precision of optical elements are obtainable, the other is to find a new method to conquer the limitation of technology. The idea of cascaded weak measurements based ultra-small phase amplification (CWMPA) offers such a method.

The idea of the CWMPA is inspired by the famous Cavendish torsion balance experiment in which the ultra-small gravitational force is amplified twice. The basis of the CWMPA is the fact that the state of pointer $|\varphi\rangle_{P}$, after post-selection, has the same form as the state of pointer $|H\rangle+e^{i\theta}|V\rangle$ with the phase signal $\theta$ introduced and the error only has an order of magnitude $o(\theta^{2})$. With the extended dimensions of the system, the amplified phase signal $\gamma$ thus can be amplified via WMPA again and again until the total error approach unacceptable level. For simplicity of illustration, we will consider a system with three dimensions and demonstrate two successive amplification of the phase signal $\theta$. 

Suppose that photons are initially prepared in the superposed path state $|\psi_{i}\rangle_{S}=a|0\rangle+b|1\rangle+c|2\rangle$ with $|a|^{2}+|b|^{2}+|c|^{2}=1$ and polarization state $|+\rangle_{P}$. Consider the control-rotation operation
\begin{equation}
\hat{U}=(|0\rangle\langle 0|+|1\rangle\langle 1|)\otimes\hat{I}+|2\rangle\langle 2|\otimes(|H\rangle\langle H|+e^{i\theta}|V\rangle\langle V|)
\end{equation}
such that the evolved state of photons becomes
\begin{equation}
|\Psi_{f}\rangle_{SP}=(a|0\rangle+b|1\rangle)\otimes|+\rangle+c|2\rangle\otimes(|H\rangle+e^{i\theta}|V\rangle)/\sqrt{2}.
\end{equation}
We first perform the post-selection in the subspace spanned by ${|1\rangle,|2\rangle}$ with post-selected path state $|3\rangle=d|1\rangle+f|2\rangle$. The composite state of photons, after post-selection, becomes (unnormalized)
\begin{equation}
\begin{split}
&|\tilde{\Psi}_{p}\rangle_{SP}=(|0\rangle\langle 0|+|3\rangle\langle 3|)|\Psi_{f}\rangle_{SP} \\
=& a|0\rangle\otimes|+\rangle+|3\rangle\otimes[(bd+cf)|H\rangle+(bd+cfe^{i\theta})|V\rangle]/\sqrt{2}
\end{split}
\end{equation}
with success probability $\mathrm{Tr}(|\tilde{\Psi}_{p}\rangle_{SP}\langle\tilde{\Psi_{p}}|)$.
Here we implicitly assume that all coefficients to be real. For the ultra-small phase $\theta\ll 1$, we have
\begin{equation}
|\tilde{\Psi}_{p}\rangle_{SP}=a|0\rangle\otimes|+\rangle+(bd+cf)|3\rangle\otimes(|H\rangle+e^{i\gamma}|V\rangle)/\sqrt{2}
\end{equation}
with
\begin{equation}
\mathrm{tan}\gamma=\dfrac{\mathrm{sin}\theta}{\mathrm{cos}\theta+bd/cf}
\end{equation}
since that $bd+cfe^{i\theta}=(bd+cf)e^{i\gamma}$ in the first order approximation with error of order $o(\theta^{2})$. When the post-selected state $|3\rangle$ is properly chosen, we can obtain the amplified phase $\gamma >\theta$. The ability of amplification is usually limited by technical imperfections in practice. In order to amplify phase signal further, we then perform post-selection in the space spanned by ${|0\rangle,|3\rangle}$ with post-selected path state $|4\rangle=m|0\rangle+n|3\rangle$. The polarization state of photons, which acts as the probe, after post-selection, becomes (unnormalized)
\begin{equation}
|\tilde{\varphi}\rangle_{P}=\langle 4|\tilde{\Psi}_{p}\rangle_{SP}=(am+qn)|H\rangle+(am+qne^{i\gamma})|V\rangle
\end{equation}
with $q=bd+cf$. As $am+qne^{i\gamma}=(am+qn)e^{i\phi}$ in the first order approximation with error of order $o(\gamma^{2})$ when $\gamma\ll 1$, the final state of the pointer reads
\begin{equation}
|\varphi\rangle_{P}=\dfrac{1}{\sqrt{2}}(|H\rangle+e^{i\phi}|V\rangle)
\end{equation}
with
\begin{equation}
\mathrm{tan}\phi=\dfrac{\mathrm{sin}\gamma}{\mathrm{cos}\gamma+am/qn}.
\end{equation}
The amplified phase signal $\phi$ can be obtained when $m,n$ are properly chosen such that $am+qn\rightarrow 0$. The total amplification factor of two successive amplifications is given by
\begin{equation}
h_{t}=h_{1}\times h_{2}=\dfrac{\gamma}{\theta}\times \dfrac{\phi}{\gamma}=\dfrac{\phi}{\theta},
\end{equation}
where $h_{1}, h_{2}$ represent the factor of amplification in the first and second amplification respectively.
Although we only demonstrate the case of two successive amplification above, it applies to the case of three or more as long as the total error of order is limited in a acceptable level. The number of successive amplification is determined by the total error $\Delta\theta_{t}$, which depends on the order of magnitude of the $\theta$ and the ability of amplification in each amplification limited by technical imperfections. Assume that the factor of amplification in each amplification has the same order of magnitude denoted by $\bar{h}$, then the approximate number of cascade amplification $N$ is given by (See Appendix for detail derivation)
\begin{equation}
N=\dfrac{\mathrm{ln}(4\epsilon\theta^{-2})/\mathrm{ln}{\bar{h}}+1}{2},
\end{equation}
where $\epsilon=\Delta\theta_{t}/\theta$ represents the precision of error that given by practical requirement. In the case of gravitational waves detection $\theta\approx 10^{-12}$, if $\bar{h}=10^{2}$ and $\epsilon=10^{-6}$ then $N=5$. This value is consistent with the intuitively estimation by $(\bar{h})^{N}=1/\theta$. In the practical situation, however, the $N$ will be limited by our technical competence of realizing cascaded amplification.

\begin{figure}[tbp]
\centering
\includegraphics[scale=0.42]{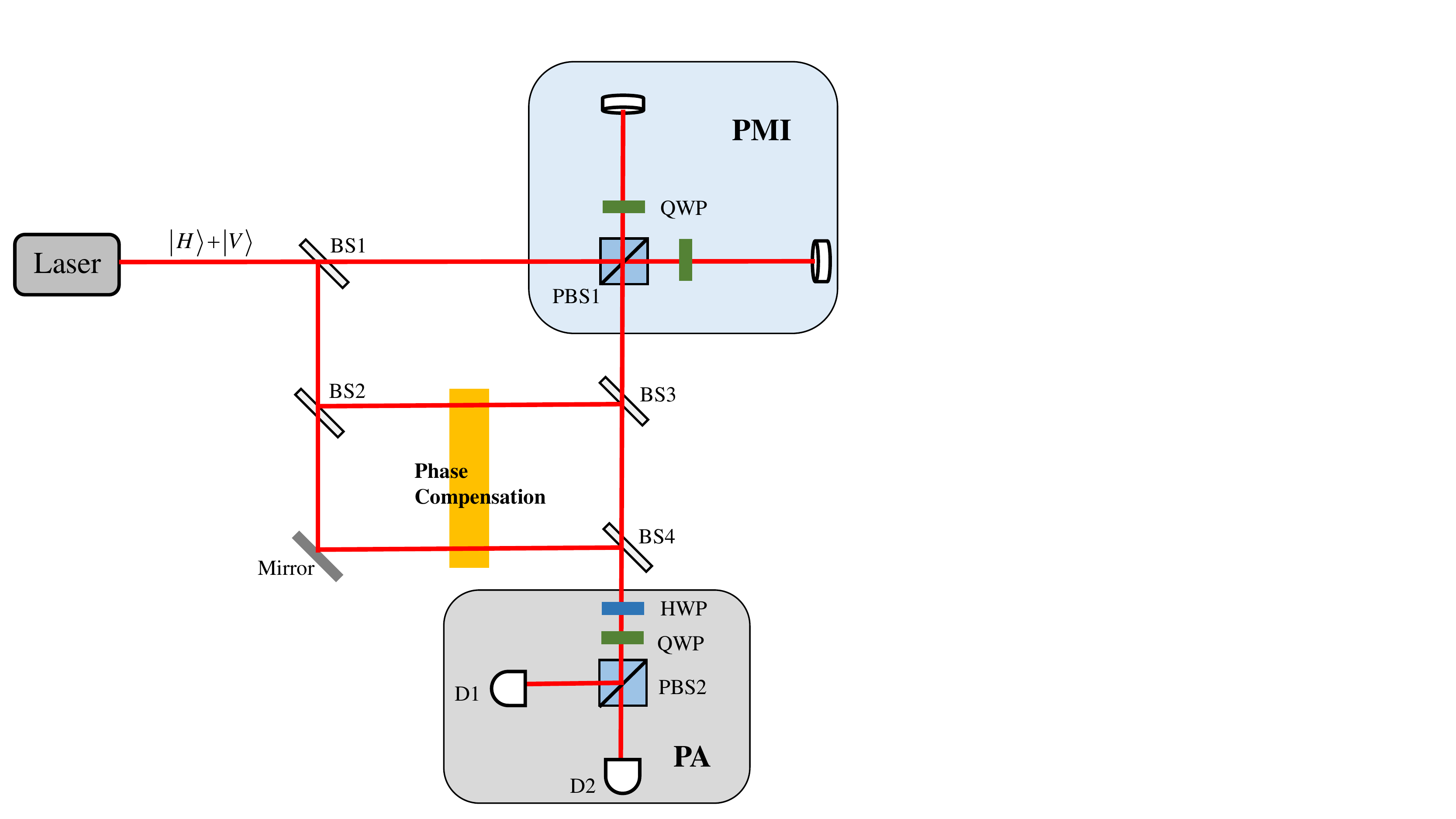}
\caption{Schematic diagram of the cascade weak measurements amplification of ultra-small phase. BS: beam splitter; PBS: polarizing beam splitter; HWP: half wave plate; QWP: quarter wave plate; D: detector; PMI: polarizing Michelson interferometer; PA: polarization analyser. The coefficient of reflection and transmission of BS1, BS2, BS3 and BS4 are $r_{1}, t_{1}; r_{2}, t_{2}; r_{3},t_{3}$ and $r_{4}, t_{4}$ respectively. The BS1 and BS2 are used for initial state preparation. The original phase signal $\theta$ is collected by PMI. The first post-selection and second post-selection are fulfilled by the BS3 and BS4 respectively. The final amplified phase signal is extracted by the PA. }
\end{figure}

Fig. 1 shows the schematic diagram of realizing two successive amplification of ultra-small phase based on weak measurements amplification. It can be easily extended to the case of more successive amplification measurement. The setup of cascaded amplification shown in Fig. 1 consists of six parts i.e., laser source, initial state preparation, signal collection, first amplification, second amplification and signal detection. The laser source with input mode cleaner produces linear polarized light in the state $|+\rangle=(|H\rangle+|V\rangle)/\sqrt{2}$ with high stability. BS1 and BS2, which have coefficients of reflection and transmission $r_{1},t_{1}$ and $r_{2},t_{2}$ respectively, realize preparation of the initial state of photons
\begin{equation}
|\Psi\rangle_{SP}=(t_{1}|up\rangle+r_{1}r_{2}|middle\rangle+r_{1}t_{2}|down\rangle)\otimes|+\rangle,
\end{equation}
where $|up\rangle,|middle\rangle,|down\rangle$ represent the path states of up arm, middle arm and down arm respectively. The photons, which transmits the BS1 in the up arm, flies to the polarizing Michelson interferometer (PMI) that used for the collection of phase signal $\theta$. The PMI consists of one polarizing beam splitter (PBS1), two quarter wave plate (QWP) that rotated at $\pi/4$ and two reflective mirrors. It outputs photons from the down port of PBS1 with the measured phase signal $\theta$ manifests itself as rotational angle of polarization state of output photons. When the input photons is in the polarization state $(|H\rangle+|V\rangle)/\sqrt{2}$, the output photons has the polarization state $(|H\rangle+e^{i\theta}|V\rangle)/\sqrt{2}$. The polarization state of photons that flies along path of middle and down arms will not be changed such that the state of photons before post-selection becomes
\begin{equation}
\begin{split}
|\Psi_{f}\rangle_{SP}&=t_{1}|up\rangle\otimes(|H\rangle+e^{i\theta}|V\rangle)/\sqrt{2} \\
&+(r_{1}r_{2}|middle\rangle+r_{1}t_{2}|down\rangle)\otimes|+\rangle.
\end{split}
\end{equation}
The first post-selection is completed by BS3 in which the post-selected photons comes out from the down port of BS3 with the path state
\begin{equation}
|\mu\rangle=t_{3}|up\rangle+r_{3}|middle\rangle,
\end{equation}
where $r_{3},t_{3}$ are coefficients of reflection and transmission of BS3. The composite state of photons, after first post-selection, reads (unnormalized)
\begin{equation}
|\tilde{\Psi}_{p}\rangle_{SP}=r_{1}t_{2}|down\rangle\otimes|+\rangle+\kappa|\mu\rangle\otimes(|H\rangle+e^{i\gamma}|V\rangle)/\sqrt{2}
\end{equation}
in the first order approximation with $\kappa=t_{1}t_{3}+r_{1}r_{2}r_{3}$ and
\begin{equation}
\mathrm{tan}\gamma=\dfrac{\mathrm{sin}\theta}{\mathrm{cos}\theta+(r_{1}r_{2}r_{3})/(t_{1}t_{3})}.
\end{equation}
The amplified phase signal $\gamma$ can be obtained by choosing $t_{3},r_{3}$ properly. Due to the technical imperfection e.g., non-ideal of BS3, the amplification factor $h_{1}=\gamma/\theta$ is usually limited and the second amplification is needed for further amplification. The second amplification, which depends on post-selection, is fulfilled by BS4 with coefficients of reflection and transmission $r_{4}$ and $t_{4}$. Similar to the case of BS3, the post-selected photons comes out from the down port of the BS4 with the path state
\begin{equation}
|\nu\rangle=t_{4}|\mu\rangle+r_{4}|down\rangle.
\end{equation}
The polarization state of the post-selected photons, in the first order approximation, becomes
\begin{equation}
|\varphi\rangle_{P}=\langle\nu|\tilde{\Psi}_{f}\rangle_{SP}=\dfrac{1}{\sqrt{2}}(|H\rangle+e^{i\phi}|V\rangle)
\end{equation}
with
\begin{equation}
\mathrm{tan}\phi=\dfrac{\mathrm{sin}\gamma}{\mathrm{cos}\gamma+(r_{1}t_{2}r_{4})/(\kappa t_{4})}.
\end{equation}
The final amplified phase signal $\phi$ can be extracted by sending the post-selected photons into the polarization analyser (PA) consists of a HWP, a QWP and PBS2. In the case that measurement basis is $\lbrace|R\rangle,|L\rangle\rbrace$ with $|R\rangle=(|H\rangle+i|V\rangle)/\sqrt{2}$ and $|L\rangle=(|H\rangle-i|V\rangle)/\sqrt{2}$, the difference intensity of the two detectors $D_{1},D_{2}$ gives
\begin{equation}
\Delta I=I_{0}\mathrm{sin}\phi=I_{0}\mathrm{sin}(h_{t}\theta)
\end{equation}
where $I_{0}$ is the intensity of post-selected light and $h_{t}$ is the total amplification factor.

In conclusion, the cascaded weak measurements amplification of a ultra-small phase is introduced here, which can be used to further improve the ability of amplification limited by technical imperfections. 
This method, which inspired by Cavendish torsion balance experiment, is based on the previous work of weak measurements based amplification of ultra-small phases \cite{hu} and can be very useful in the practical high precision measurement e.g., the detection of gravitational waves when technical issues must be taken into considerations.

This work was supported by the National Natural Science Foundation of China (No. 61275122, No. 11674306 and No. 61590932), the Strategic Priority Research Program (B) of the Chinese Academy of Sciences (No. XDB01030200) and National key R$\&$D program (No. 2016YFA0301300 and No. 2016YFA0301700).

\section{Appendix}
In this appendix, we give a detail derivation of cascaded number $N$ i.e, Eq. (16) in main body.

In the discussion of WMPA, we have ignored the influence of higher order term $o(\theta^{2})$ to the phase signal estimation because that $\theta\ll 1$. However, in the case of CWMPA, the error will also be cascaded amplified such that the total error set a limitation on the cascaded number of weak measurements amplification. Assume that error amplification is a linear process, the total error of estimation after $N$ cascaded amplification will be determined by recursion formula
\begin{equation}
 \Delta\theta_{N}=\bar{h}\Delta\theta_{N-1}+\Delta\vartheta_{N},
\end{equation} 
where $\Delta\varphi_{N}$ is the error of estimation due to higher order term in the process of $N$th amplification.
Since that $\Delta\theta_{1}=\Delta\vartheta_{1}$, it is necessary to calculate $\Delta\vartheta_{1}$ i.e., the error of estimation in the first amplification. In the WMPA, the pointer's state after post-selection in the first order approximation is given by Eq. (4). In order to extract amplified phase $\gamma$, we perform measurement of observable $\hat{\sigma}_{R}\equiv |R\rangle\langle R|-|L\rangle\langle L|$ on the pointer, which gives $_{P}\langle\varphi|\hat{\sigma}_{R}|\varphi\rangle_{P}=\mathrm{sin}\gamma$. However, the actual measurement result should be $_{P}\langle\tilde{\varphi}|\hat{\sigma}_{R}|\tilde{\varphi}\rangle_{P}=[(\mathrm{cos}\chi+\mathrm{sin}\chi)/\sqrt{(\mathrm{cos}\chi+\mathrm{sin}\chi)^{2}-\mathrm{sin}(2\chi)\cdot\epsilon}]\cdot\mathrm{sin}\gamma$ with $\epsilon=1-\mathrm{cos}\theta$. The higher order term of $\theta$ result in the error $\Delta\gamma$ of $\gamma$ estimation and thus the error $\Delta\vartheta$ of $\theta$ estimation. The error $\Delta\gamma$ is given by
$\mathrm{sin}(\gamma+\Delta\gamma)=_{P}\langle\tilde{\varphi}|\hat{\sigma}_{R}|\tilde{\varphi}\rangle_{P}$. In the first order approximation $\mathrm{sin}(\gamma+\Delta\gamma)=\mathrm{sin}\gamma+\mathrm{cos}\gamma\cdot\Delta\gamma$ and thus we have
\begin{equation}
\Delta\gamma=[\dfrac{\mathrm{cos}\chi+\mathrm{sin}\chi}{(\mathrm{cos}\chi+\mathrm{sin}\chi)^{2}-\mathrm{sin}(2\chi)\cdot\epsilon}-1]\cdot\mathrm{tan}\gamma.
\end{equation}
Since that $\mathrm{tan}\gamma=\mathrm{sin}\theta/(\mathrm{cos}\theta+\mathrm{cot}\chi)$, $\chi=-(\pi/4+\delta)$ with $\delta\ll 1$ and $(1-\epsilon)^{-1/2}=1+\epsilon/2$ in the first order approximation, the $\Delta\gamma$ is given as
$\Delta\gamma=(\bar{h}^{2}\theta^{3})/4$, where $\epsilon=\theta^{2}/2$ in the first order approximation and $\bar{h}\approx 1/\delta$. The estimation error $\Delta\vartheta_{1}$ is thus obtained as
\begin{equation}
\Delta\vartheta_{1}=\Delta\gamma/\bar{h}=\dfrac{1}{4}\bar{h}\theta^{3}.
\end{equation}
The repeated use of recursion formula gives
\begin{equation}
\Delta\theta_{N}=\dfrac{1}{4}\bar{h}\theta^{3}\sum\bar{h}^{2(N-1)}.
\end{equation}
Defining $f(N)=\sum\bar{h}^{2(N-1)}$, then $(\bar{h}^{2}-1)f(N)=\bar{h}^{2N}-1$ results in $f(N)=(\bar{h}^{2N}-1)/(\bar{h}^{2}-1)\approx\bar{h}^{2(N-1)}$, which gives $\Delta\theta_{N}=\bar{h}^{2N-1}\theta^{3}/4$. If we denote the precision of error as $\varepsilon$ ($\varepsilon\ll 1$), then $\Delta\theta_{N}=\varepsilon\theta$ gives
\begin{equation}
\bar{h}^{2N-1}=4\varepsilon\theta^{-2}.
\end{equation}
The final cascaded number $N$ of weak measurements amplification is thus given by Eq. (16) in main body.


\begin{thebibliography}{99}

\bibitem{hu} Meng-Jun Hu and Yong-Sheng Zhang, Gravitational Wave Detection via Weak Measurements, arXiv:1707.00886 [quant-ph].












\bibitem{wva1} Y. Aharonov, D. Z. Albert, and L. Vaidman, How the result of a measurement of a component of the spin of a spin-1/2 particle can turn out to be 100, Phys. Rev. Lett. {\bf 60}, 1351 (1988).








\bibitem{wva2} O. Hosten and P. Kwiat, Observation of the Spin Hall Effect of Light via Weak Measurements, Science {\bf 319}, 787 (2008).

\bibitem{wva3} P. B. Dixon, D. J. Starling, A. N. Jordan, and J. C. Howell, Ultrasensitive Beam Deflection Measurement via Interferometric Weak Value Amplification, Phys. Rev. Lett. {\bf 102}, 173601 (2009).

\bibitem{wva4} X. Y. Xu, Y. Kedem, K, Sun, L. Vaidman, C. F. Li, and G. C. Guo, Phase Estimation with Weak Measurement Using a White Light Source, Phys. Rev. Lett. {\bf 111}, 033604 (2013).


\bibitem{wva5} J. Dressel, M. Malik, F. M. Miatto, A. N. Jordan, and R. W. Boyd, {\it Colloquium:} Understanding quantum weak values: Basics and applications, Rev. Mod. Phys. {\bf 86}, 307 (2014).

\bibitem{bird} J. C. Bird, F. Liang, B. H. Solheim, and G. G. Shepherd, A polarizing Michelson interferometer for measuring thermospheric winds, Means. Sci. Technol. {\bf 6}, 1368-1378 (1995).
























\end{thebibliography}
\end{document}